\documentclass[aoas,nameyear,rotating,dvips]{arximspdf}

\doi{10.1214/10-AOAS350}
\volume{4}
\issue{2}
\pubyear{2010}
\firstpage{663}
\lastpage{686}

\makeatletter
\DeclareMathAlphabet\mathcaligr{OMS}{cmsy}{m}{n}
\renewcommand{\mathcal}{\mathcaligr}
\renewcommand{\citep}{\citet}
\renewcommand{\cite}{\citet}
\renewcommand{\parallel}{\Vert}
\renewcommand{\epsilon}{\varepsilon}

\makeatother

\begin{document}
\begin{frontmatter}

\title{Sparse regulatory networks}
\runtitle{Sparse regulatory networks}

\begin{aug}
\author[A]{\fnms{Gareth M.} \snm{James}\corref{}\thanksref{t1}\ead[label=e1]{gareth@usc.edu}},
\author[B]{\fnms{Chiara} \snm{Sabatti}\thanksref{t2}\ead[label=e2]{sabatti@stanford.edu}},
\author[C]{\fnms{Nengfeng} \snm{Zhou}\ead[label=e3]{iambosszhou@gmail.com}}
\and
\author[C]{\fnms{Ji}~\snm{Zhu}\thanksref{t3}\ead[label=e4]{jizhu@umich.edu}}
\runauthor{James, Sabatti, Zhou and Zhu}
\affiliation{University of Southern California, Stanford University,
University of Michigan and University of Michigan}
\thankstext{t1}{Supported in part by
NSF Grants DMS-07-05312 and DMS-09-06784.}
\thankstext{t2}{Supported in part by NIH/NIGMS Grant GM053275-14.}
\thankstext{t3}{Supported in part by NSF Grants DMS-07-05532 and DMS-07-48389.}
\address[A]{G. M. James\\
Marshall School
of Business\\
University of Southern California\\
Hoffman Hall 512\\
Los Angeles, California 90089\\
USA\\
\printead{e1}} 
\address[B]{C. Sabatti\\
Department of Health Research\\
\quad and Policy\\
Stanford University\\
HRP Redwood Building\\
Stanford, California 94305-5405\\
USA\\
\printead{e2}}
\address[C]{N. Zhou\\J. Zhu\\
Department of Statistics\\
University of Michigan\\
1085 South University Ave.\\
Ann Arbor, Michigan 48109-1107\\
USA\\
\printead{e3}\\
\hphantom{E-mail: }\printead*{e4}}
\end{aug}

\received{\smonth{1} \syear{2009}}
\revised{\smonth{3} \syear{2010}}

%
\begin{abstract}
In many organisms the expression levels of each gene are controlled by
the activation levels of known ``Transcription Factors'' (TF). A problem
of considerable interest is that of estimating the ``Transcription
Regulation Networks'' (TRN) relating the TFs and genes. While the
expression levels of genes can be observed, the activation levels of the
corresponding TFs are usually unknown, greatly increasing the difficulty
of the problem. Based on previous experimental work, it is often the case
that partial information about the TRN is available. For example,
certain TFs may be known to regulate a given gene or in other cases a
connection may be predicted with a certain probability. In general, the
biology of the problem indicates there will be very few connections
between TFs and genes. Several methods have been proposed for
estimating TRNs. However, they all suffer from problems such as
unrealistic assumptions about prior knowledge of the network structure
or computational limitations. We propose a new approach that can
directly utilize prior information about the network structure in
conjunction with observed gene expression data to estimate the TRN. Our
approach uses $L_1$ penalties on the network to ensure a sparse
structure. This has the advantage of being computationally efficient as
well as making many fewer assumptions about the network structure. We use
our methodology to construct the TRN for E. coli and show that the
estimate is biologically sensible and compares favorably with previous
estimates.
\end{abstract}

%

\end{frontmatter}

\section{Introduction}

Recent progress in genomic technology allows scientists to gather vast and
detailed information on DNA sequences, their variability, the timing and
modality of their translation into proteins, and their abundance and
interacting partners. The fields of system and computational biology have
been redefined by the scale and resolution of these data sets and the
necessity to interpret this data deluge. One theme that has clearly
emerged is the importance of discovering, modeling and exploiting
interactions among different biological molecules. In some cases, these
interactions can be measured directly, in others they can be inferred from
data on the interacting partners. In this context, reconstructing
networks, analyzing their behavior and modeling their characteristics have
become fundamental problems in computational biology.

\begin{figure}

\includegraphics{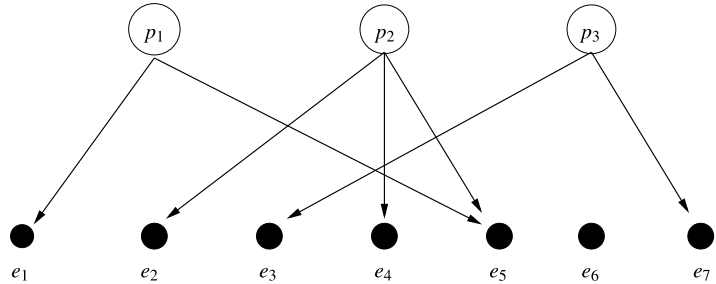}

\caption{A general network with $L=3$ transcription factors and $n=7$
genes.}\label{network}
\end{figure}

Depending on the type of biological process considered, and the type of
data available, different network structures and graph
properties are relevant. In this work we focus on one type of bipartite
network that has been used to model transcription regulation, among other
processes, and is illustrated in Figure~\ref{network}. One
distinguishes input nodes
($p_1, p_2, p_3$ in Figure~\ref{network}) and output nodes ($e_1,
\ldots, e_7$ in
Figure~\ref{network}); directed edges connect input nodes to one or more
output nodes and indicate control. Furthermore, we can associate a numerical
value with each edge, which indicates the nature and strength of
the control.

%

Bipartite networks such as the one illustrated in Figure~\ref{network}
have been
successfully used to describe and analyze transcription regulation [see,
e.g., \cite{liao1}]. Transcription is the initial step of the
process whereby the information stored in genes is used by the cell
to assemble proteins. To adapt to different cell functions and different
environmental conditions, only a small number of the genes in the DNA are
transcribed at any given time. Understanding this selective process is the
first step toward understanding how the information statically coded in
DNA dynamically governs all the cell life. One critical role in the
regulation of this process is played by transcription factors. These
molecules bind in the promoter region of the genes, facilitating or making
it impossible for the transcription machinery to access the relevant
portion of the DNA. To respond to different environments, transcription
factors have multiple chemical configurations, typically existing both in
``active'' and ``inactive'' forms. Their binding affinity to the DNA
regulatory regions varies depending on the particular chemical
configuration, allowing for a dynamic regulation of transcription.
Depending on the complexity of the organism at hand, the total number of
Transcription Factors (TF) varies, as well as the number of TF
participating in the regulation of each gene. In bipartite networks
such as
the one in Figure~\ref{network}, input nodes can be taken to represent
the variable
concentrations in active form of transcription factors, and output
nodes as
the transcript amounts of different genes. An edge connecting a TF to a
gene indicates that the TF participates in the control of the gene
transcription. As usual, mathematical stylization only captures a
simplified version of reality. Bipartite graphs overlook some
specific mechanisms of transcription regulation, such as
self-regulation of TF
expression or feed-back loops connecting genes to transcription factors.
Despite these limitations, networks such as the one in Figure~\ref
{network} provide a
useful representation of a substantial share of the biological
process.

Researchers interested in reconstructing transcription regulation have at
their disposal a variety of measurement types, which in turn motivate
diverse estimation strategies. The data set that motivated the development
of our methodology consisted of measurements of gene transcription levels
for E. coli, obtained from a collection of $35$ gene expression arrays.
These experiments, relatively cheap and fairly common, allow one to
quantify transcription amounts for all the genes in the E. coli genome,
under diverse cell conditions.
While our data consists of measurements on the output nodes, that is,
the gene expression levels, we also have access to some information
on the topology of the network: DNA sequence analysis or ChIP--chip
experiments can be used to evaluate the likelihood of each possible edge.
However, we have no direct measurements of the input nodes, that is, the
concentrations of active form of the TFs. While, in theory, it is
possible to obtain these measurements, they are extremely expensive and
are typically unavailable. Changes in transcription of TF are measured
with gene expression arrays, 
but mRNA levels of transcription factors seldom correlate with changes in
the concentration of their active form. The latter, in fact, are most often
driven by changes in TF expression level only in response to the cell
inner clock (i.e., in development, or in different phases of the cell
cycle). We are interested in studying the cellular response to external
stimuli and this is most frequently mediated by post-translational
modifications of the TF. For these reasons, we are going to consider the
concentrations of active forms of the TF as unobserved.

Our E. coli data consist of spotted array experiments with two dyes, which
measure the changes in expression from a baseline level for the queried
genes (taking the logarithm of the ratio of intensities, typically
reported as raw data). These percentage changes can be related linearly to
variations in the concentrations of active form of transcription factors,
as documented in \cite{liao1}.
Coupling this linearity assumption, with
the bipartite network structure, we model the
log-transformed expressions of gene $i$ in experiment $t$, $e_{it}$, as
\[
e_{it}=\sum_{j=1}^La_{ij}p_{jt}+\epsilon_{it}, \qquad  i = 1, \ldots, n,
  t= 1,\ldots, T,
\]
where $n$, $L$ and $T$ respectively denote the
number of genes, TFs and experiments, $a_{ij}$ represents the control
strength of transcription factor $j$ on gene $i$, $p_{jt}$ the
concentration of the active form of transcription factor $j$ in experiment
$t$, and $\epsilon_{it}$ captures \textit{i.i.d.} measurement errors and biological
variability. A value of $a_{ij}=0$ indicates that there is no network
connection or, equivalently, no relationship, between gene $i$ and TF $j$,
while nonzero values imply that changes in the TF affect the gene's
expression level. It is convenient to formulate the model in matrix
notation,
%
\begin{equation}
\label{basemodel}
E=AP+\epsilon,
\end{equation}
where $E$ is an $n\times T$ matrix of $e_{it}$'s, $A$ is an $n\times L$
matrix of $a_{ij}$'s and $P$ is an $L \times T$ matrix of $p_{jt}$'s. $A$
and $P$ are both unknown quantities.

Model (\ref{basemodel}), derived from the bipartite regulatory network
and linearity assumption, is a very familiar one to statisticians and a
number of its variants have been applied to the
study of gene expression and other data. The first attempts utilized
dimension reduction techniques such as principal component analysis (PCA)
or singular value decomposition [\cite{alter1}]. Using this approach, a
unique solution to simultaneously estimate the $p_j$'s and the strength of
the network connections is obtained by assuming orthogonality of the
$p_j$'s---an assumption that does not have biological motivations. Some
variants of PCA, that aim to produce more interpretable results, have also
been studied. For example, \citeauthor{lee1nnmf} (\citeyear{lee1nnmf,lee2nnmf}) developed
nonnegative matrix factorization (NNMF) where the elements of $A$ and $P$
are all constrained to be positive. However, for our data we would expect
both positive and negative control strengths, so it does not seem reasonable
to enforce the elements of $A$ to be positive. An interesting development
is the use of Independent Component Analysis [\cite{lee}], where the
orthogonality assumption is substituted by stochastic independence. These
models can be quite effective in providing a dimensionality reduction, but
the resulting $p$'s often lack interpretability.

\cite{west1} treats (\ref{basemodel}) as a factor model and uses a
Bayesian approach to reduce the dimension of expression data, paying
particular attention to the development of sparse
models, in order to achieve a biologically realistic representation. When
the gene expression data refers to a series of experiments in a meaningful
order (temporal, by degree of exposure, etc.), model (\ref{basemodel}) can
be considered as the emission component of a state space model, where
hidden states can be meaningfully connected to transcription factors
[\cite{beal}, \cite{li}, \cite{sanguinetti}]. Depending on the amount of knowledge
assumed on the $A$ matrix, state space models can deal with networks of
different size and complexity.

Values of the factors, $P$, that are clearly interpretable as changes in
concentration of the active form of transcription factors together with
the identifiability of model (\ref{basemodel}) can be achieved by imposing
restrictions on $A$ that reflect available knowledge on the topology of
the network. \cite{liao1} assumes the entire network structure known
a priori and gives conditions for identifiability of~$A$ and~$P$ based on
the pattern of zeros in $A$, reflecting the natural sparsity of the
system. A simple iterative least squares procedure is proposed for
estimation, and the bootstrap used to asses variability. This approach
has two substantial limitations. First, it assumes that the entire network
structure is known, while, in practice, it is most common for only
parts of
the structure to have been thoroughly studied. Second, not all known
transcription networks satisfy the identifiability conditions. A number
of subsequent contributions have addressed some of these limitations.
\cite{tran} introduces other, more general, identifiability conditions;
\cite{yu} proposes an alternative estimation procedure for the factor
model; \cite{brynildsen} explores the effect of inaccurate specification
of the network structure; \cite{chang} proposes a faster algorithm.
\cite{pournara1} provides an informed review of the use of factor models
for regulatory networks, surveying both different identifiability
strategies and computational approaches.

Particularly relevant to the present paper is the work of \cite{sabatti1},
which removes both limitations of the \cite{liao1} method by using a
Bayesian approach. The authors obtain a prior probability on the network
structure using sequence analysis, and then use a Gibbs sampler to produce
posterior estimates of the TRN. In theory, this approach can be applied to
any network structure, even when only part of the structure is known.
However, a significant limitation is that the computational effort
required to implement the Gibbs sampler grows exponentially with the
number of potential connections between a particular gene and the
transcription factors. As a result, one is forced to choose a prior on the
network where the probability of most edges is set to zero, thereby fixing
a priori a large portion of the topology. While sparsity in the
connections is biologically reasonable, it would obviously be more
desirable to allow the gene expression data to directly identify the
connections.

To overcome these limitations, in this paper we take a somewhat different
approach that builds in the same advantages as the Bayesian method in
terms of utilizing partial network information and working for any
structure. However, our approach is more computationally efficient, which
allows increased flexibility in determining the final network topology. We
treat the estimation of both the connection strengths, $A$, and the
transcription factors concentrations, $P$, as a variable selection
problem. In this context, our data has an extremely large number of
variables, that is, potential connections, but is sparse in terms of the
number of ``true'' variables, that is, connections that actually exist. There
have recently been important methodological innovations for this type of
variable selection problem. A~number of these methods involve the use of
an $L_1$ penalty on the regression coefficients which has the effect of
performing automatic variable selection. A few examples include the Lasso
[\cite{tibshirani3}], SCAD [\cite{fan2}], the Elastic Net [\cite{zouelastic}],
the adaptive Lasso [\cite{zou1}], the Dantzig selector [\cite{candes1}],
the Relaxed Lasso [\cite{meinshausen2}], VISA [\cite{james12}] and the
Double Dantzig [\cite{james10}]. The most well known of these approaches
is the Lasso, which performs variable selection by imposing an $L_1$
penalty on the regression coefficients. In analogy with the Lasso, our
method also utilizes $L_1$ penalties on the connection strengths, $A$, as
well as the transcription factor concentrations, $P$. This allows us to
automatically produce a sparse network structure, which incorporates the
prior information. We show that, given the same prior network, our
approach produces similar results to the Bayesian formulation, but is
considerably more computationally efficient.
This in turn allows us to reconstruct regulatory networks using less
precise prior information.

%
\begin{figure}

\includegraphics{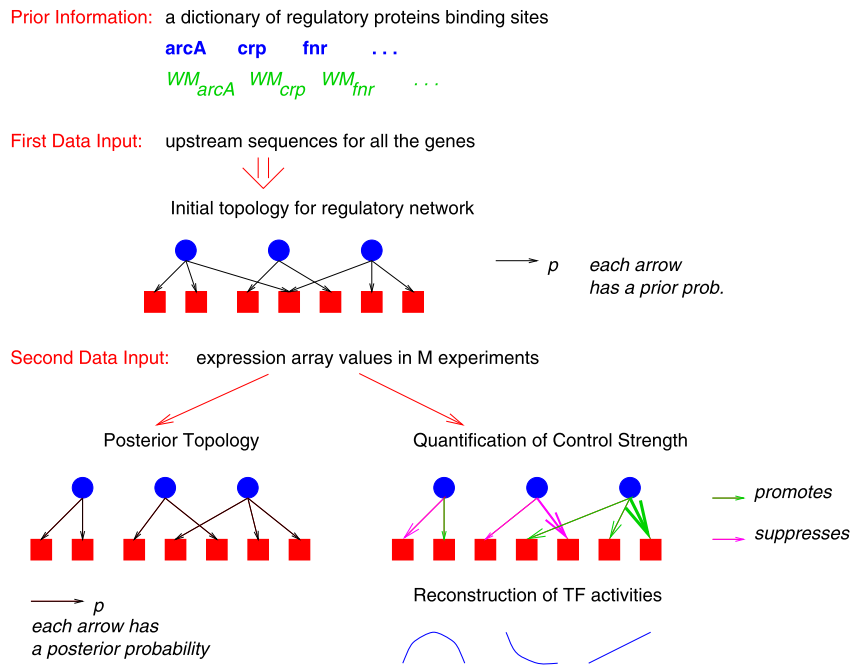}

\caption{Transcription network reconstruction
integrating DNA sequence and gene expression information. Blue circles
represent regulatory proteins and red squares genes. An arrow connecting
a circle to a square indicates that the transcription factor controls
the expression of the gene. When different colors are used in depicting
these arrows, they signify a different qualitative effect of the TF on
genes (repressor or enhancer). Finally, varying arrow thickness signifies
different control strengths.}\label{flowchart}
\end{figure}

Figure~\ref{flowchart} gives a schematic illustration of our approach.
First, we identify a group of transcription factors that are believed to
regulate the gene expression levels. Second, we compute an initial
topology for the network using both documented experimental evidence, as
well as an analysis of the DNA sequence upstream of a given gene.
Finally, we use the initial topology, as well as the gene
expression levels from multiple experiments, as inputs to our
$L_1$ penalized regression approach to produce an updated final network
topology, a quantification of the connection strengths and an estimation
of the transcription factor levels.

The paper is structured as follows. In Section~\ref{datasec} we
provide a
detailed description of the data that we are analyzing and the
available prior information.
Section~\ref{methodsec} develops the methodological approach
we use to fit the transcription regulation network. Our analysis of the E.
coli data is presented in Section~\ref{casesec}. We also include a
comparison with the results using the Bayesian approach in
\cite{sabatti1}. A simulation study where we compare our approach with two
other possible methods is provided in Section~\ref{simsec},
followed by a discussion in Section~\ref{discsec}.\looseness=1

\section{Data and prior information on network structure}
\label{datasec}

The data set that motivated the development of our methodology included
$35$ microarray experiments of \textit{Escherichia coli} that were
either publicly available
or were carried out in the laboratory of Professor James C. Liao at UCLA.
The experiments consisted of Tryptophan timecourse data (1--12)
[\citep{khodursky1}], glucose acetate transition data (13--19)
[\citep{oh1}, \citep{oh2}], UV exposure data (20--24) [\citep{courcelle1}] and a protein
overexpression timecourse data set (25--35) [\citep{oh3}].
In all cases, gene expression arrays allow us to monitor the cellular
response to external stimuli: as noted in the introduction, this is
mediated by changes in concentration of active forms of the transcription
factors. Current knowledge alerts us that the TrpR regulon should be
activated in the Tryptophan timecourse data, the LexA regulon should be
activated in the UV experiments, and the RpoH regulon in the protein
overexpression data. To provide the reader with a clearer picture of the
underlying biology, we detail the case of Tryptophan starvation and UV
exposure. The Trp operon encodes enzymes necessary for synthesis of the
amino acid tryptophan; it is suppressed by TrpR, which can bind to the DNA
only in the presence of Tryptophan. When Tryptophan is depleted, TrpR stops
acting as a suppressor, and the Trp operon is transcribed. Treating
Escherichia coli with radiation produces some damage, which, in turn,
induces a number of cellular responses, aiming at counteracting it. One
well-known response is called SOS and is controlled by the RecA and LexA
proteins. Typically, LexA represses SOS genes. When single-stranded DNA,
produced as a result of radiation damage, is present in the cell, it binds
to the RecA protein, activating its protease function; the activated
RecA cuts
the LexA protein, which can no longer act as a repressor, and the SOS
genes are induced. Note that both TrpR and LexA auto-regulate, but
post-translational modifications play a dominant role in changing their
concentration of active form in response to external stimuli.

To reduce spurious
effects due to the inhomogeneity of the data collection, we standardized
the values of each experiment, so that the mean across all genes in each
experiment was zero and the variance one. Merging these different data sets
resulted in expression measurements on $1433$ genes across $35$ experiments.

We also were able to identify partial information about the network
structure connecting the transcription factors and genes. 
We first identified a set of transcription factors that previous
literature suggested were important in this system: this resulted in 37
transcription factors. Our bipartite network structure can be represented
using the $n \times L$ matrix $A$ of control strengths where $n= 1433$ is
the number of genes under consideration and $L=37$ is the number of
transcription factors. Note that the fact that we consider more
transcription factors (37) than experiments (35) makes it impossible to
analyze this network structure using the NCA framework presented by
\cite{liao1}.

The element $a_{ij}$ is nonzero if TF $j$ regulates
gene $i$, and zero otherwise. For a number of well-studied TF, experimental
data is available that clearly indicates their binding in the upstream
region of regulated genes (in other words, $a_{ij}\ne0$).
However, for many of the elements of $A$, only partial information is
available. To summarize the prior evidence on the network structure, we
introduce $\pi_{ij}=P(a_{ij}\ne0)$. If there is documented experimental
evidence of a binding site for transcription factor $j$ in the promoter
region of gene $i$, we set $\pi_{ij}=1$. We assign values to the
remaining elements of $\pi$ using an analysis of the DNA sequence upstream
of the studied genes. We use available information on the characteristics
of the DNA sequence motif recognized by the TF to inform the sequence
analysis, carried out with \textit{Vocabulon} [\cite{sabatti2}]. Vocabulon
produces an estimated probability that TF $j$ controls gene $i$ which we
used as an initial value for~$\pi_{ij}$. This algorithm is particularly
well suited for this genomewide investigation, but other methodologies
could also be applied. We hence identify all the putative binding sites
for these transcription factors in the portion of the genome sequence that
is likely to have a regulatory function.

Two qualifications are in order. First, resorting to Vocabulon and
sequence analysis is only but one venue to gather knowledge on the
network structure. In particular, it is worth noting that results from
ChIP--Chip experiments are an important source of information that could
be used for this purpose (see \cite{boulesteix} and \cite{sun} for a
detailed study of these data). Second, the degree of sparsity of the
initial network can be substantially varied, as documented in
Section~\ref{resultssec}.
Indeed, one can use different thresholds to decide when a binding site is
detected; moreover, putative sites may have a varying degree of certainty
that could be reflected in the choice of $\pi_{ij}$. However, we have
found that the most important issue is assuring that $\pi$ does not play
an excessive part in the fitting procedure so that the expression data
can make a significant contribution to the final estimated TRN. In
Section~\ref{priorsec} we discuss a shrinkage approach that ensures the
prior is not overly informative.

\section{Methodology}
\label{methodsec}

\subsection{A preliminary approach}
\label{prelimsec}
A natural way to extend the Lasso procedure to fit our model
(\ref{basemodel}) is to minimize, over $A$ and $P$, the penalized squared
loss function:
%
\begin{equation}
\label{eq01}
\parallel E - AP
\parallel_2^2 + \lambda_1
\parallel A
\parallel_1 + \lambda_2 \parallel P \parallel_1,
\end{equation}
where $\lambda_1$ and $\lambda_2$ are two tuning parameters and
$\Vert \cdot\Vert _1$ is the sum of the absolute values of the given matrix. Note
that $\parallel\cdot\parallel_2^2$ corresponds to the sum of squares
of all
components of the corresponding matrix with any missing values ignored.
While this objective function appears to require the selection of two
tuning parameters, (\ref{eq01}) can be reformulated as
\[
\parallel E - A^*P^*
\parallel_2^2 + \lambda_1\lambda_2
\parallel A^*
\parallel_1 + \parallel P^* \parallel_1,
\]
where $A^*=A/\lambda_2$ and $P^*=\lambda_2P$. Hence, it is clear that a
single tuning parameter suffices and $A$ and $P$ can be computed as the
minimizers of
%
\begin{equation}
\label{eq02}
\parallel E - AP
\parallel_2^2 + \lambda
\parallel A
\parallel_1 + \parallel P \parallel_1.
\end{equation}
Optimizing (\ref{eq02}) for different values of $\lambda$ controls the
level of sparsity of the estimates for $A$ and $P$.

A simple iterative algorithm can be used to solve (\ref{eq02}), namely:
\begin{itemize}
\item Step 1: Choose initial values for $A$ and $P$ denoted by
$A^{(0)}$ and
$P^{(0)}$. Let $k = 1$.
\item Step 2: Fix $A=A^{(k-1)}$, find the $P = P^{(k)}$ minimizing $
\parallel E - A^{(k-1)}P
\parallel_2^2 + \parallel P \parallel_1 $.
\item Step 3: Fix $P= P^{(k)}$, find the $A = A^{(k)}
$ minimizing $
\parallel E - AP^{(k)}
\parallel_2^2 + \lambda\parallel A \parallel_1 $.
\item Step 4: If $ \parallel P^{(k)} - P^{(k-1)} \parallel$ or $
\parallel A^{(k)} - A^{(k-1)}
\parallel$ are large, let $k\leftarrow k+1$ and return to Step 2.
\end{itemize}
Steps 2 and 3 in this algorithm can be easily achieved using a standard
application of the LARS algorithm [\cite{efron3}] used for fitting the Lasso.

\subsection{Incorporating the prior information}
\label{priorinfosec}
The fitting procedure outlined in the previous section is simple to
implement and often quite effective. It can be utilized in situations
where no prior information is available about the network structure
because minimizing (\ref{eq02}) is, a priori, equally likely to cause any
particular element of $A$ to be zero, or not to be zero.

However, in practice, for our data, we know that many elements of $A$ must
be zero, that is, where $\pi_{ij}=0$, and others cannot be zero, that
is, where
$\pi_{ij}=1$. Of the remaining elements, some are highly likely to be zero,
while others are most likely nonzero, depending on their $\pi_{ij}$.
Hence, it is important that our fitting procedure directly takes the prior
information into account. This limitation is removed by minimizing
(\ref{finalfit1}),
%
\begin{equation}
\label{finalfit1}
\Vert E-AP\Vert _2^2 - \lambda_1 \sum_{ij} \log(\pi_{ij}) |a_{ij}| +
\lambda_2
\Vert A\Vert _2^2 + \Vert P\Vert _1.
\end{equation}
The key changes between (\ref{eq02}) and (\ref{finalfit1}) are the
addition of $-\log(\pi_{ij})$ and a square of $L_2$ norm penalty on
$A$. The
incorporation of the prior information has several effects on the fit.
First, $a_{ij}$ is automatically set to zero if $\pi_{ij}=0$. Second,
$a_{ij}$ cannot be set to zero if $\pi_{ij}=1$. Finally, $a_{ij}$'s for
which the corresponding $\pi_{ij}$ is small are likely to be set to zero,
while those for which $\pi_{ij}$ is large are unlikely to be set to zero.
Optimizing (\ref{finalfit1}) is achieved using a similar iterative
approach to that used for (\ref{eq02}):%
\begin{itemize}
\item Step 1: Choose initial values for $A$ and $P$ denoted by
$A^{(0)}$ and
$P^{(0)}$. Let $k = 1$.
\item Step 2: Fix $A=A^{(k-1)}$, find the $P = P^{(k)}$ minimizing $
\parallel E - A^{(k-1)}P
\parallel_2^2 + \parallel P \parallel_1 $.
\item Step 3: Fix $P= P^{(k)}$, find the $A = A^{(k)}
$ minimizing $
\parallel E - AP^{(k)}\parallel_2^2
- \lambda_1 \sum_{ij} \log(\pi_{ij}) |a_{ij}|+ \lambda_2
\Vert A\Vert _2$.
\item Step 4: If $ \parallel P^{(k)} - P^{(k-1)} \parallel$ or $
\parallel A^{(k)} - A^{(k-1)}
\parallel$ are large, let $k\leftarrow k+1$ and return to Step 2.
\end{itemize}
Step 2 can be again be implemented using the LARS algorithm. Step 3
utilizes the shooting algorithm [\cite{fu1}, \cite{friedman1}].

Equation (\ref{finalfit1}) treats all elements of $P$ equally.
However, in
practice, there is often a grouping structure in the experiments or,
correspondingly, the columns of~$P$. For example, in the E. coli data
columns $1$ through $12$ of $P$ correspond to the Tryptophan timecourse
experiments, while columns $13$ through $19$ represent the glucose acetate
transition experiments. To examine any possible advantages from modeling
these natural groupings, we implemented a second fitting procedure. Let~$\mathcal{G}_k$ be the index of the experiments in the $k$th group
assuming all the experiments are divided into $K$ groups. Then our second
approach involved minimizing,
%
\begin{equation}
\label{finalfit2}
\Vert E-AP\Vert _2^2 - \lambda_1 \sum_{ij} \log(\pi_{ij}) |a_{ij}| +
\lambda_2
\Vert A\Vert _2^2 + \Vert P\Vert _2,
\end{equation}
where $\Vert  P\Vert _2 = \sum_{j=1}^{L} \sum_{k=1}^{K} \sqrt{\sum_{t \in
\mathcal{G}_k} p_{jt}^2 }$. Replacing $\Vert  P \Vert _1$ with $\Vert  P\Vert _2$ has
the effect of forcing the $p_{jt}$'s within the same group to either all
be zero or all nonzero. In other words, either all of the experiments or
none of the experiments within a group are selected. Minimizing
(\ref{finalfit2}) uses the same algorithm as for (\ref{finalfit1}) except
that in Step 2 the shooting algorithm is used rather than LARS. We show
results from both methods. To differentiate between the two approaches, we
call (\ref{finalfit1}) the ``ungrouped'' method and (\ref{finalfit2}) the
``grouped'' approach.

Both equations (\ref{finalfit1}) and (\ref{finalfit2}) bare some
resemblance to the penalized matrix decomposition (PMD) approach
[\cite{witten1}]. PMD is a general method for decomposing a matrix, $E$,
into matrices, $A$ and $P$. As with our method, PMD imposes various
penalties on the components of $A$ and $P$ to ensure a sparse, and hence
more interpretable, structure. However, the decomposition it produces is
more similar to standard PCA because it does not attempt to incorporate
any prior information, instead imposing orthogonality constraints on~$A$
and~$P$.

Our methodology does not make any explicit assumptions about the
distribution of the error terms, $\epsilon_{it}$. However, it is worth
noting that if we model the error terms as i.i.d. Gaussian random variables,
then, with the variance term fixed, the likelihood function associated
with this model is inversely proportional to $\| E -AP\|_2^2$. Hence, equations
(\ref{eq02}), (\ref{finalfit1}) and (\ref{finalfit2}) can all be
viewed as
approaches to maximize the penalized likelihood; the only difference between
methods being in the form of the penalty function.

\subsection{Adjusting the prior}
\label{priorsec}

The grouped and ungrouped methods both assume a known prior, $\pi
_{ij}$. In
reality, the prior must itself be estimated. In some situations this
can be
done with a reasonable level of accuracy. However, in other instances the
estimated prior may suggest a much higher level of certainty than it is
reasonable to assume. For instance, sequence analysis algorithms, such as
Vocabulon, tend to produce many probability estimates that are very close
to either $0$ or $1$. In reality, a sequence analysis can usually only
provide an indication as to whether a connection exists between a
particular TF and gene, so a probability closer to $0.5$ may be more
appropriate.

To account for this potential bias in the prior estimates, we adjust the
initial prior using the following equation:
%
\begin{equation}
\label{shrinkage}
\tilde\pi_{ij} =
\cases{ 0, &\quad  $\pi_{ij}=0$,\cr
(1-\alpha)\times\pi_{ij} +\alpha\times0.5, &\quad  $0<\pi_{ij}<1$,\cr
1, &\quad  $\pi_{ij}=1$,
}
\end{equation}
where $\tilde\pi_{ij}$ represents the
adjusted prior. The shrinkage parameter, $\alpha$, represents the
level of
confidence in the initial prior. A value of $\alpha=0$ corresponds to a
high level of confidence in the estimated prior. In this situation no
shrinkage is performed and the prior is left unchanged. However, values of
$\alpha$ close to $1$ indicate much lower confidence. Here the estimated
probabilities that are strictly between $0$ and $1$ are shrunk toward
$0.5$, corresponding to an uninformative prior. As documented in
Section~\ref{casesec}, we experimented with various different values for
$\alpha$.

\subsection{Normalizing the estimators}
\label{normsec}

The use of penalties on $A$ and $P$ will generally allow us to produce
unique estimates for the parameters up to an indeterminacy in the signs of
$A$ and $P$, that is, one can obtain identical results by flipping the
sign on
the $j$th column of $A$ and the $j$th row of $P$. There are a number of
potential approaches to deal with the sign. \cite{sabatti1} defined two
new quantities that are independent from rescaling and changes of signs
and have interesting biological interpretations:
\[
\tilde p_{jt} = \frac{\sum_i a_{ij} p_{jt}}{\sum_i 1(a_{ij}\ne0)}
\quad \mbox{and} \quad  \tilde a_{ij} = \frac{\sum_t a_{ij} p_{jt}}{T}.
\]
$\tilde{p}_{jt}$ is the average effect of each transcription factor on the
genes it regulates (regulon expression), and $\tilde{a}_{ij}$ is the
average control strength over all experiments. These quantities are
directly related to the expression values of genes in a regulon. We have
opted to use $\tilde p_{jt}$ and $\tilde a_{ij}$ to report our results.
This also has the advantage of allowing easy comparison with the analysis
of \cite{sabatti1}.

Providing general conditions on the prior for identifiability is complex
and beyond the scope of this paper. In general, the more zero, or close to
zero, elements there are in $\pi$, the more likely the model is to be
identifiable. Alternatively, it is easy to show that as $\min\pi_{ij}
\rightarrow1$ the model will become unidentifiable. The results in
\cite{liao1} and \cite{tran} can be used to provide sufficient conditions
for identifiability when the prior has enough elements close to zero.
These conditions, which we provide in the \hyperref[app]{Appendix}, are similar to those
given in \cite{Banderson1} for identifiability of factor models. The
\hyperref[app]{Appendix} also contains details of an empirical study we conducted using
multiple randomized starting points for our algorithm. The results
suggested that there were no identifiability problems for the E. coli
data.

\section{Case study}
\label{casesec}

In this section we give a detailed examination of the results from
applying the grouped and ungrouped methods to the E. coli data.
Section~\ref{firststructure} outlines the construction of our initial
network structure, while Section~\ref{tuningsec} discusses our procedure
for choosing the tuning parameters. The main results are provided in
Section~\ref{resultssec}. Finally, Section~\ref{relaxsec} gives the
results from a sensitivity analysis performed by adjusting the sparsity
level on the initial network structure. All the results reported in
Section~\ref{casesec} represent the optimal fit, in terms of the final
objective values, based on ten randomized initial values of $A$ and $P$.

\subsection{The initial network structure}
\label{firststructure}

The first step in constructing the transcription regulation network is to
develop an initial guess for $\pi$, that is, the probability
distribution of
the network structure. As discussed in Section~\ref{datasec}, $\pi$ was
computed using various sources. Where there was experimental evidence
of a
link between transcription factor $j$ and gene $i$ we set $\pi_{ij}=1$.
For the remaining elements we used the \textit{Vocabulon} [\cite{sabatti2}]
algorithm to estimate $\pi_{ij}$. We then adjusted the prior estimates
using the shrinkage approach, Equation (\ref{shrinkage}), which required
selecting a value for the shrinkage parameter, $\alpha$. We experimented
with four different values for $\alpha$; $0, 0.3, 0.65$ and $1$. In most
instances it did not have a significant effect on the final results,
suggesting our method is robust to changes in the nonzero values of the
prior. For our final analysis we opted to use $\alpha=1$ because this
produced the weakest prior which gave the gene expression data the best
opportunity to determine the final network structure. Note, our initial
prior estimates contained a number of values corresponding to exactly $0$
or $1$, so even after performing the shrinkage step our new prior still
contained enough information to ensure an identifiable solution.
In addition, this
approach produced similar priors to those used in \cite{sabatti1} which
allowed us to directly compare the two sets of results. With the Bayesian
approach of \cite{sabatti1}, this high level of sparsity in the network
structure was necessary for computational reasons. However, using our
Lasso based methodology, this level of sparsity is not required. Hence, in
Section~\ref{relaxsec} we examine how our results change as we reduce the
level of sparsity in the initial structure.

By merging the potential binding sites with the known sites from the
literature, and with the expression data, we obtained a set of $1433$
genes, potentially regulated by at least one of $37$ transcription factors
and on which expression measurements were available (missing values in the
array data were allowed). Our estimate for~$\pi$ suggested a great
deal of
sparsity with only $2073$ nonzero entries, $291$ of which corresponded to
$\pi_{ij}=1$ and the remaining $1782$ to $\pi_{ij}=0.5$.
In addition, $14$ of the transcription factors were expected to regulate
$20$ or fewer genes and $34$ of the $37$ TFs were expected to
regulate at most $120$ genes. The notable exception was CRP, which
potentially regulated over $500$ genes. It is worth noting that without
adopting our penalized regression framework, we would not be able to study
this transcription network, simply because the number of experiments
($35$) is smaller than the number of TF considered ($37$): the use of
penalty terms regularizes the problem.

\subsection{Selecting the tuning parameters}
\label{tuningsec}
The first step in estimating $A$ and $P$ requires the selection of the
tuning parameters, $\lambda_1$ and $\lambda_2$. These could be chosen
subjectively but we experimented with several more objective automated
approaches. We first attempted to select the tuning parameters
corresponding to the lowest values of BIC or AIC. However, BIC produced
models that were biologically too sparse, that is, the number of zero entries
in $A$ was too large. It appears that the $\log(n)$ factor used by BIC is
too large if one uses the number of nonmissing values in the $E$ matrix
as ``$n$'' ($n=40$,$000$) because they are not really independent. Conversely,
AIC resulted in networks being selected that had too many connections.

%
\begin{figure}[b]

\includegraphics{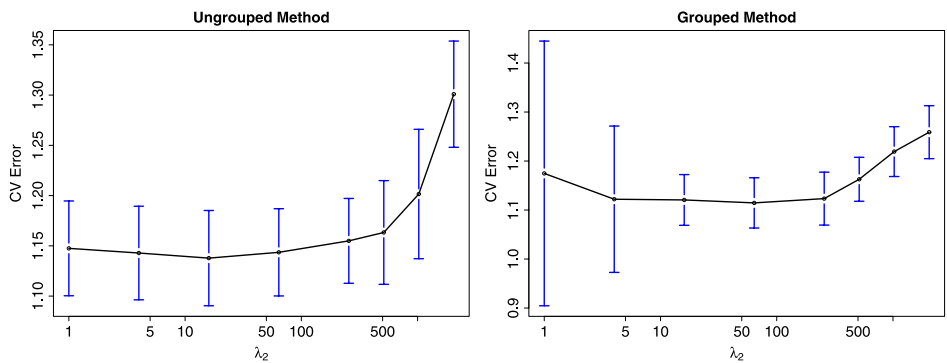}

\caption{Cross validated error rates as a
function of $\lambda_2$ for the ungrouped and grouped methods. The blue
vertical lines indicate variability in the cross validated
error.}\label{tuningparams}
\end{figure}

Instead we opted to use a two stage approach. We first computed the
``relaxed'' cross validated error over a grid of $\lambda_1$'s and
$\lambda_2$'s and selected the tuning parameters corresponding to the
minimum. It is well known that cross validation can perform poorly on
model selection problems involving $L_1$ penalties [\cite{meinshausen1}].
This is mainly a result of shrinkage in the coefficient estimates. A
common approach to reduce the shrinkage problem in the Lasso involves
replacing the nonzero coefficients with their corresponding least squares
estimates. Our relaxed cross validation approach works in a similar way.
For each combination of $\lambda_1$ and $\lambda_2$, we first use
equations (\ref{finalfit1}) and (\ref{finalfit2}) to identify initial
estimates for $A$ and $P$. We then fix $P$ and the zero elements of $A$ and
use ``least squares'' to estimate the nonzero elements of $A$. The cross
validated errors are then computed based on these ``un-shrunk'' estimates
for $A$. We have found that this approach allows us to select sparser
network structures than those from using standard cross validation.
Figure~\ref{tuningparams} shows the cross validated error rates for
different values of $\lambda_2$ with $\lambda_1=64$. For the grouped
method the minimum was achieved with
$\lambda_1=\lambda_2=64$, while the ungrouped minimum was achieved with
$\lambda_1=64$ and $\lambda_2=16$.\looseness=-1

Second, we used a parametric bootstrap analysis to determine
whether there was significant evidence that an element in $A$ was
nonzero. We ran our method on $100$ bootstrap samples, each created
by
first computing the residuals $\hat e = E - \hat A \hat P$, resampling
$\hat e$, and then generating the bootstrap sample $E^{(b)} = \hat A
\hat P
+ \hat e^{(b)}$. For each element
of $A$, we computed a corresponding $p$-value based on the $100$ bootstrap
results, thus, we had approximately $2000$ $p$-values. Since this constituted
a significant multiple testing problem, we used False Discovery Rate (FDR)
methods to set a cutoff such that the FDR was no more than $0.05$.
Elements in $A$ with $p$-values smaller than the cutoff were left as is
while the remainder were set to zero. All the results that follow are
based on this bootstrap analysis.

\subsection{Results}
\label{resultssec}

\begin{figure}[b]

\includegraphics{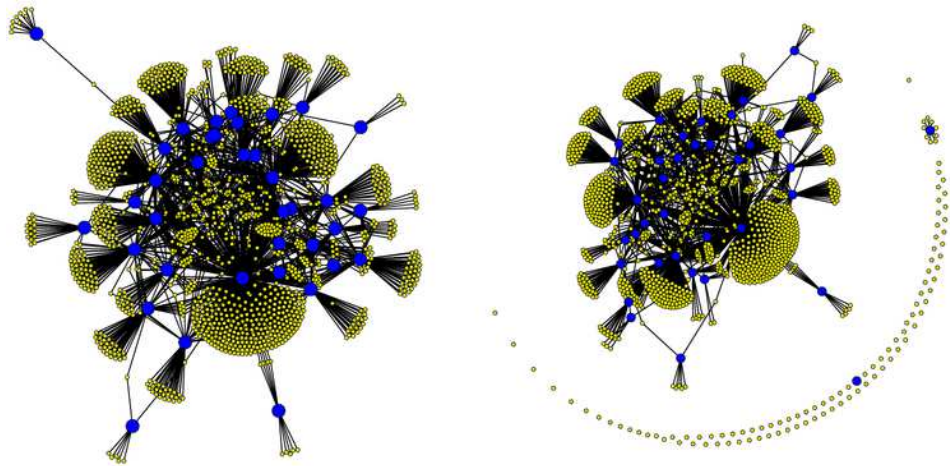}

\caption{Prior network (left) and posterior estimate
produced using the ungrouped method (right). The large blue circles
correspond to the $37$ transcription factors while the yellow circles
represent the $1433$ genes. The lines joining blue and yellow circles
indicate network connections.}\label{nets}
\end{figure}

The results from our analysis of the $35$ experiments suggested that a
significant portion of the potential binding sites should be discarded.
Now $18$ TFs
were expected to regulate $20$ or fewer genes and $26$ of the
$37$ TFs were expected to regulate at most $50$ genes. Even CRP went from
over $500$ potential binding sites in the prior to fewer than $500$ in the
posterior. The posterior estimate for $A$ contained $1766$ nonzero
entries, approximately a $15\%$ reduction in the number of connections
compared to our prior guess for the network. Figure~\ref{nets} provides
graphical representations for the prior and posterior networks. Notice
that in the posterior estimate there are many fewer connections and, as a
result, there are numerous genes and one TF that are no longer
connected to
the rest of the network, suggesting there is no evidence that these
particular genes are regulated by any of the $37$ TFs we examined. The
fact that one of the TFs is not connected to the network is likely due to
it not being activated in any of the experiments considered, so that there
is no detectable correlation in expression among the group of genes that
it regulates.

%

\cite{sabatti1} discuss several possible reasons for the changes between
the initial and final network structure. In brief, Vocabulon works
entirely using the sequence information. Hence, it is quite possible
for a
portion of the E.~coli genome sequence to look just like a binding site
for a TF, resulting in a high probability as estimated by Vocabulon, when
in reality it is not used by the protein in question. In addition,
Vocabulon searches for binding sites in the regulatory region of each gene
by inspecting $600$ base pairs upstream of the start codon which often causes
Vocabulon to investigate the same region for multiple genes. If a binding
site is located in such a sequence portion, it will be recorded for all of
the genes whose ``transcription region'' covers it.

%
\begin{figure}[b]
\centering
\begin{tabular}{@{}c@{}}

\includegraphics{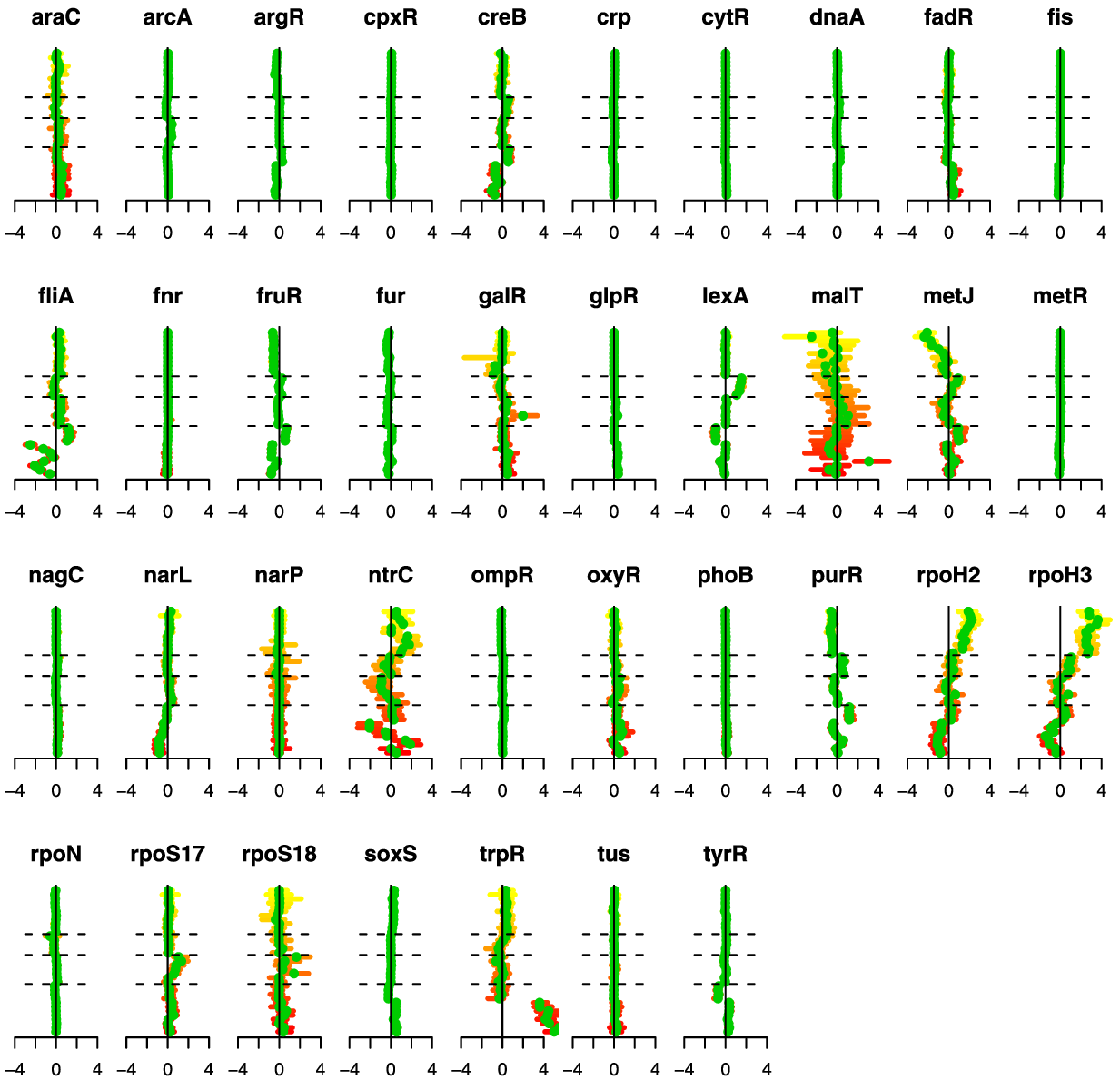}
\\
\footnotesize{(a)}
\end{tabular}
\caption{\textup{(a)} Ungrouped and \textup{(b)} grouped methods.
Each plot
corresponds to the experiments for one transcription factor. Experiments
are organized along the vertical axis, from bottom to top, with dashed
lines separating the experiment groups. Green dots indicate the
estimates for $\tilde p_{jt}$ and the horizontal bars provide bootstrap
confidence intervals.}\label{L1P}\vspace*{-5pt}
\end{figure}
\setcounter{figure}{4}
\begin{figure}
\centering
\begin{tabular}{@{}c@{}}

\includegraphics{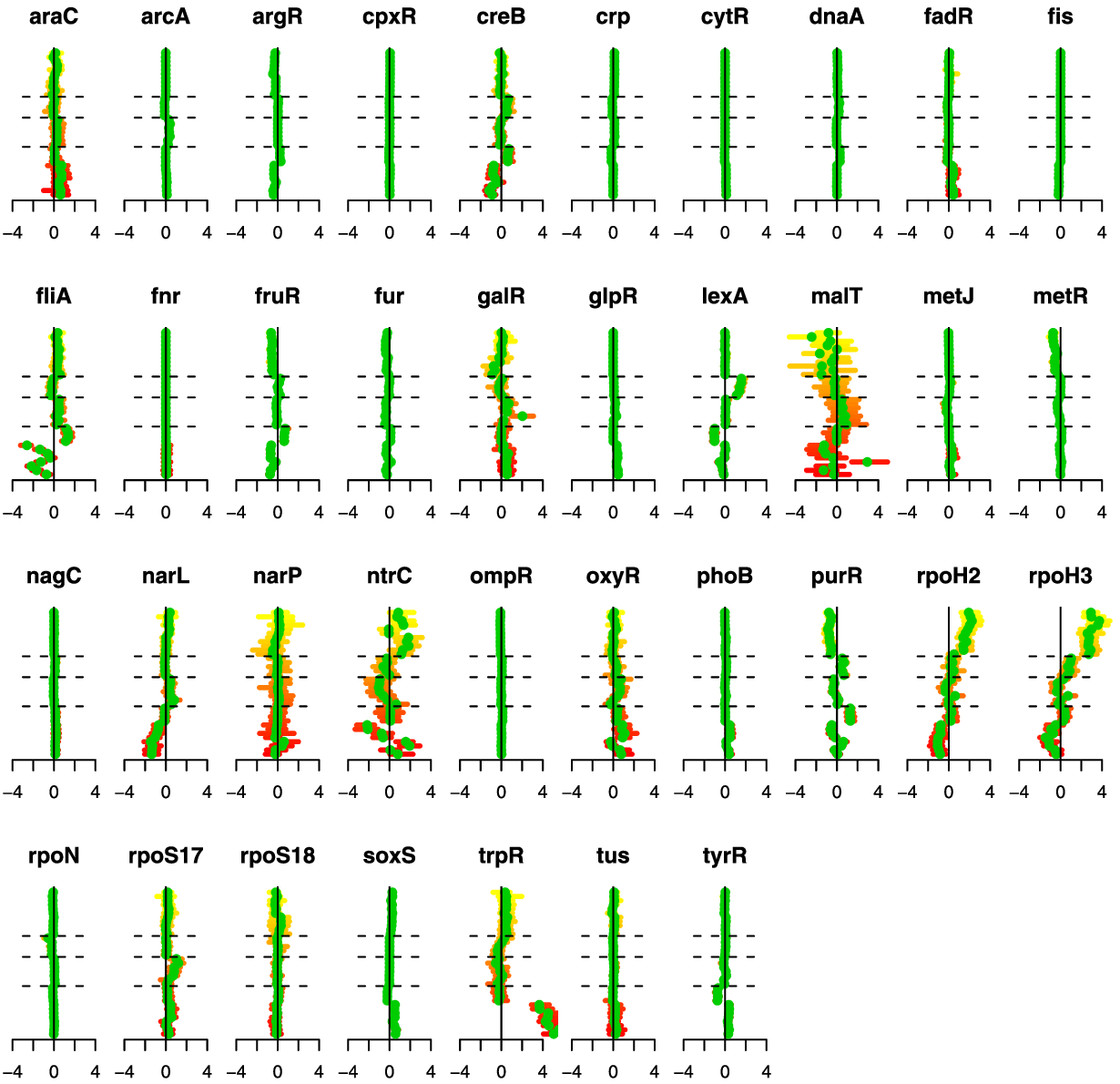}
\\
\footnotesize{(b)}
\end{tabular}
\caption{(Continued).}
\end{figure}

Figure~\ref{L1P} illustrates the estimated transcription factor activation
levels using both the ungrouped and grouped methods. We have several ways
to validate  these results. First, we note that the estimated activation
levels show very strong similarities to the results of \cite{sabatti1}.
Both their results and ours show the following characteristics. First,
there are a number of transcription factors that are not activated in any
of the experiments. Focusing on the regulons that are activated in some of
the experiments, we note that our method produces results that correspond
to the underlying biology.
For example, the first 8 experiments [\citep{khodursky1}]---represented in the lower portion of the displays from bottom up---are two
4-point time courses of tryptophan starvation. The absence of tryptophan
induces the de-repression of the genes regulated by trpR.
Correspondingly, our results indicate a clear increase in expression for
trpR. In arrays 9--12, the cells were provided with extra tryptophan.
Hence, for these experiments we would expect lowered expression. Our
results show a negative effect, though the magnitude is small.
Additionally, the argR and fliA regulons can be seen to move in the
opposite direction to trpR, which corresponds to what has been documented
in the literature [\citep{khodursky1}].

Experiments 20--24, which correspond to the results between the second and
third horizontal dashed lines, are a comparison of wild type E. coli
cells with cells that were irradiated with ultraviolet light, which
results in DNA damage. Notice that lexA
appears to be activated in
these experiments, as one would predict since many of the DNA
damaged-genes are known to be regularly repressed by lexA
[\cite{courcelle1}]. Finally, ntrC, purR, rpoH2 and rpoH3 all
show activations in the protein overexpression data, the final $11$
experiments. In particular,
notice that rpoH2 and rpoH3 present the same profile across all
experiments. This provides further validation of our procedure since these
two really represent the same protein, and are listed separately because
they correspond to two different types of binding sites of the TF.
Overall, these results conform to the known biology, but also suggest some
additional areas for exploration.

The main differences between our results and those of \cite{sabatti1}
are that
our penalties on $P$ tend to generate more exact zero estimates than the
Bayesian approach, providing somewhat easier interpretation. The grouped
and ungrouped results are also similar, but the grouped method tends to
produce slightly more sparsity in $P$, for example, in metJ and rpoS18.

%
\begin{figure}[b]

\includegraphics{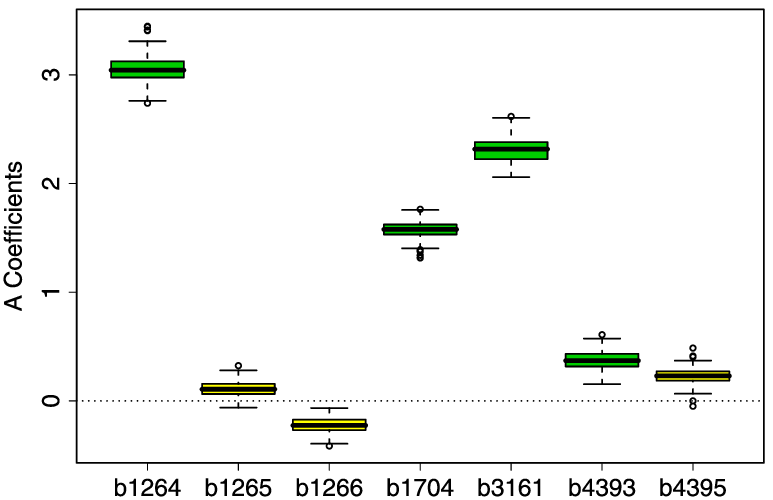}

\caption{Boxplots of the bootstrap estimates for
$\tilde a$ for seven different genes.}\label{L1Genes}
\end{figure}

Next, we examine the estimates for $A$. Since a number of TF's showed no
activation in these experiments, we would not expect to be able to
accurately estimate their control strengths on the genes. Hence, we will
concentrate our analysis here on trpR because this was the most strongly
activated TF. Figure~\ref{L1Genes} presents our estimates of $\tilde{a}$
for seven genes associated with the trpR. Each boxplot illustrates the
$100$ bootstrap estimates of $\tilde{a}$ for a particular gene. The first
three boxplots correspond to genes b1264, b1265, b1266. The b-numbers,
that identify the genes, roughly correspond to their genomic location, so
it is clear that the genes are adjacent to each other. Gene b1264 is known
to be regulated by trpR, so it's $\pi_{ij}$ was set to $1$. The other two
genes were chosen by Vocabulon as potential candidates because the binding
site for b1264 was also in the search regions for b1265 and b1266, that is,
these were cases of the overlapping regulatory regions described
previously. While Vocabulon was unable to determine whether a connection
existed between b1265, b1266 and trpR, using our approach, we can see that,
while~$\tilde{a}$ for b1264 is large, the estimates for b1265 and b1266
are essentially zero. These results show that the expression levels of
b1264 correlate well with those of the other genes, but those for b1265
and b1266 do not. Thus, it is possible to use our model to rule out the
regulation of two genes by trpR that are within a reasonable distance from
a trpR real binding site. Among the remaining four genes, b1704, b3161 and
b4393 are all known to be regulated by trpR. Correspondingly, they all
have moderate to large estimated activation strengths. b4395 again has an
overlapping regulatory region to b4393. The results suggest this is not
regulated by trpR.

\subsection{Relaxing zero coefficients}
\label{relaxsec}

The results from Section~\ref{resultssec} use the same relatively sparse
initial network structure as that of \cite{sabatti1}. Recall the
structure we have assumed so far contained only three possible values for
$\pi$, that is, $\pi_{ij}=0,$ $\pi_{ij}=0.5$ or $\pi_{ij}=1$. All
connections
with $\pi_{ij}=0$ are forced to remain at zero whatever the gene
expression data may suggest. However, as discussed previously, our
methodology is able to handle far less sparse structures. Hence, we next
investigated the sensitivity of our results to the initial structure by
randomly adjusting certain TF-gene connections. In particular, we randomly
selected $200$ of the connections where $\pi_{ij}=0$ and reset them to
$\pi_{ij}=0.5$. We also reset all connections where $\pi_{ij}=1$ to
$\pi_{ij}=0.5$ so that all connections were treated equivalently. We then
reran the ungrouped and grouped methods using the new values for $\pi$.

%
\begin{figure}

\includegraphics{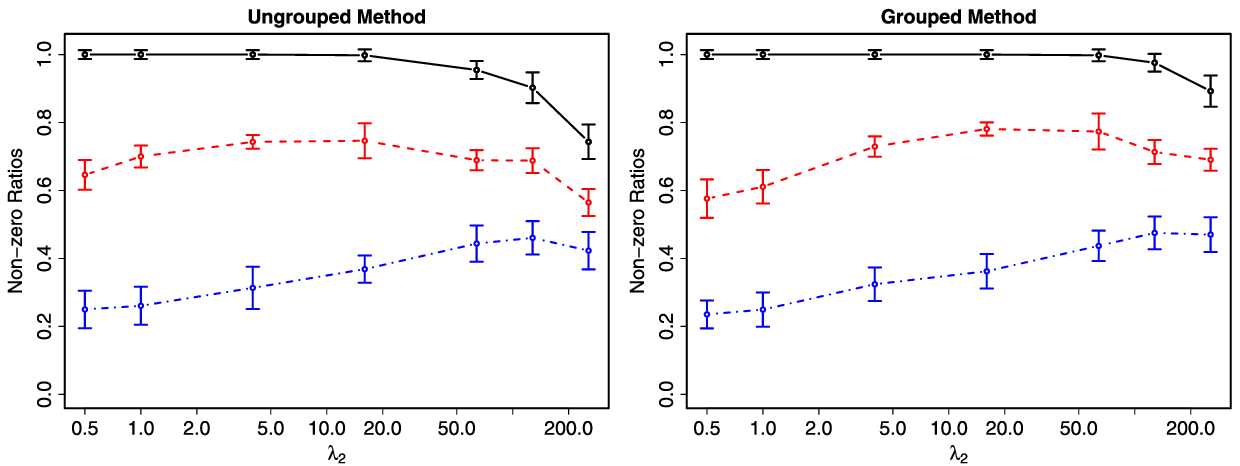}

\caption{Fraction of nonzero $\tilde a_{ij}$'s as a
function of $\lambda_2$ for the ungrouped and grouped methods. The black
solid line corresponds to those connections where there was documented
evidence of a relationship, the red dashed line to where the Vocabulon
algorithm suggested there was a relationship and the blue dash--dot line
to where there was no evidence of a relationship.}\label{nonzero}\vspace*{-1pt}
\end{figure}

Figure~\ref{nonzero} provides plots of the resulting fractions of nonzero
estimates for $\tilde a_{ij}$, as a function of $\lambda_2$ with
$\lambda_1$ set to $64$.
A clear pattern emerges with the fraction of nonzeros where there was
documented evidence very high (black solid line). Somewhat lower is the
fraction of nonzeros for the connections suggested by Vocabulon (red
dashed line). Finally, the lowest level of nonzeros is exhibited where
there was no significant evidence of a connection (blue dash--dot line).
These results are comforting because they suggest that our methodology is
able to differentiate between the clear, possible and unlikely connections
even when $\pi_{ij}$ is equal for all three groups. In addition, there
appears to be evidence that the Vocabulon algorithm is doing a good job of
separating potential from unlikely connections. Finally, these results
illustrate that, unlike the Bayesian approach, it is quite computationally
feasible for our methodology to work on relatively dense initial network
structures.

\section{Simulation study}
\label{simsec}

After fitting the E. coli data we conducted a simulation study to assess
how well our methodology could be expected to reconstruct transcription
regulation networks with characteristics similar to those for our data
set. We compared our method with two other possible approaches: the
penalized matrix decomposition (PMD) method of \cite{witten1} and the
Bayesian factor analysis model (BFM) of \cite{west1}.

The estimated matrices, $\hat A$ and $\hat{P}$, and the prior probability
estimates, $\pi_{ij}$, from Section~\ref{casesec} were used as the
starting point for generating the gene expression levels. In
particular, we
first let $\tilde{A} = \hat{A} + \varepsilon_A $, $\tilde{P} = \hat
{P} +
\varepsilon_P $, where $\varepsilon_{A_{ij}} \sim s_A \times N(0,
\sigma^2(\hat{A}) )$ and $\varepsilon_{P_{ij}} \sim s_P \times N(0,
\sigma^2(\hat{P_i}) )$ are noise terms. Depending on the simulation run,
$s_A$ was set to either $0.2$ or $0.4$, while $s_P$ was set to either $0.1$
or $0.3$. Next, all elements of $\tilde{A}$ corresponding to $\pi
_{ij} =
0$ were set to zero. In addition, among elements where $\pi_{ij} = 0.5$,
we randomly set $\rho$ of the $\tilde{A}$'s to zero where $\rho$ was set
to either $60\%$ or $80\%$. The expression levels were then generated
using
\[
E = \tilde{A} \tilde{P} + s_N \times
\tilde{\Gamma},
\]
where
$\tilde{\Gamma}$ is a matrix of error terms with $\tilde{\Gamma
}_{ij}\sim
N(0, 1 )$ and $s_N$ was set to either $0.2$ or $0.4$. We produced one
simulation run for each combination of $s_A, s_P, \rho$, and $s_N$,
resulting in a total of $16$ simulations.

%
\begin{figure}

\includegraphics{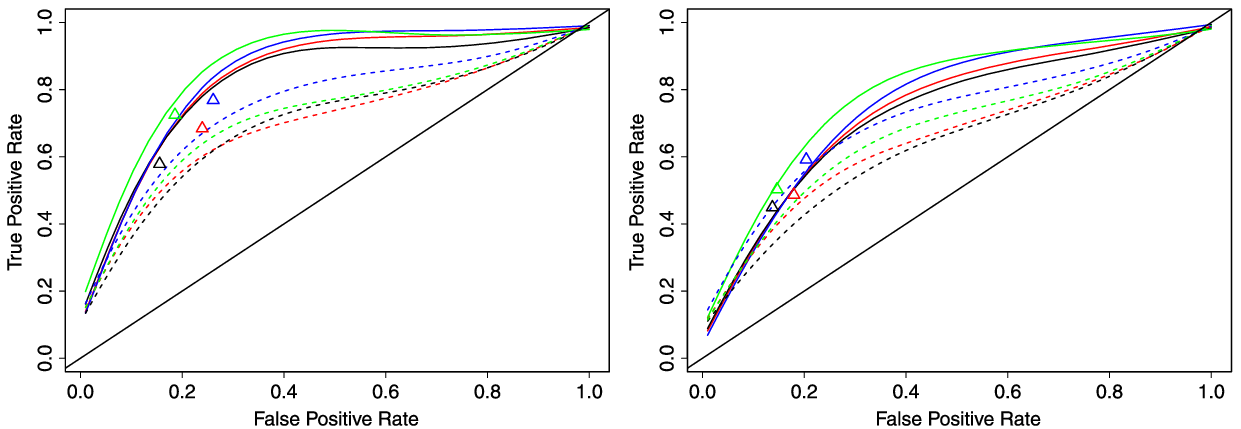}

\caption{Simulation results. Solid lines
correspond to the ungrouped approach, dashed lines to PMD and triangles
to BFM. Red: $\rho=0.6, s_A=0.2$. Black: $\rho=0.8, s_A=0.2$. Blue:
$\rho=0.6, s_A=0.4$. Green: $\rho=0.8, s_A=0.4$. Left plot: low noise
scenario, $s_N=0.2$. Right plot: high noise scenario,
$s_N=0.4$.}\label{simplot}
\end{figure}

For each simulation run we generated a new data set, implemented the
grouped and ungrouped methods, as well as the PMD method, using different
possible tuning parameters to estimate $A$ and $P$, and computed the
corresponding False Positive Rates (FPR) and the True Positive Rates
(TPR). The FPR is defined as the fraction of estimated nonzero
coefficients, $a_{ij}$, among all elements of $\tilde A$ where $\tilde
a_{ij}=0$ and $\pi_{ij}=0.5$. The TPR is defined as the fraction of
estimated nonzero coefficients, $a_{ij}$, among all elements of $\tilde
A$ where $\tilde a_{ij}\ne0$ and $\pi_{ij}=0.5$. The BFM approach turned
out to run extremely slowly, taking many hours for just a single tuning
parameter. Hence, it was only feasible to implement this method for one
set of tuning parameters. For our method, since we have prior information,
we can match the columns of the estimated $A$ with the true $A$ in order
to compute the sensitivity and specificity etc., but for both PMD and BFM,
there is no automatic alignment. In order to ensure a fair
comparison, we used a sequential alignment approach to match the
columns of
the estimated and true $A$. We first matched each column of the estimated
$A$ with each column of the true $A$ and linked the pair that matched
best. Then we removed the pair and repeated the process until all columns
were aligned.\looseness=-1

Figure~\ref{simplot} provides a summary of the results from running the
ungrouped, PMD and BFM approaches on the eight simulations corresponding
to $s_P=0.1$. The results from the grouped method and for $s_P=0.3$ were
similar and hence are not repeated here. Each curve corresponds to the FPR
vs TPR for one simulation run using different tuning parameters. The
results suggest that our method achieves a reasonable level of accuracy
for this data. For example, with $s_N=0.2$ we produce an $80\%$ TPR at the
expense of a $20\%$ FPR. To lower the FPR to $10\%$ decreases the TPR to
approximately $60\%$. Even with $s_N=0.4$, a relatively high noise level,
we can achieve a $60\%$ TPR at the expense of a $20\%$ FPR. The PMD method
performs relatively worse, for example, producing only a $60\%$ TPR at the
expense of a $20\%$ FPR with $s_N=0.2$. Assessing BFM is more difficult,
given that we were only able to observe its performance at a few points.
It appears to outperform PMD and produce results close to our ungrouped
method. However, BFM does not seem to be practical on large data sets like
our E. coli data given the time required to produce a single fit, without
even attempting to select tuning parameters. These results show that
indeed there is an advantage to including prior information when
available.\looseness=-1

\section{Discussion}
\label{discsec}

We have introduced a new methodology for estimating the parameters of
model (\ref{basemodel}) associated with a bipartite network, as
illustrated in Figure~\ref{network}. Our approach is based on
introducing $L_1$
penalties to the regression framework, and using prior information about
the network structure.

We have focused on the application of this model to reconstruction of the
E.~coli transcription network, as this allows easy comparison with
previously proposed models. Our approach has the advantage, over the work
of \cite{liao1} and \cite{sabatti1}, that it does not require assuming
prior knowledge of a large fraction of the network. When we utilize the
same prior structure as used in \cite{sabatti1}, we get similar, and
biologically sensible, results. However, by relaxing the prior assumptions
on the sparsity of the network structure, we gain additional insights such
as independent validation, both of the experimentally derived network
connections and also the connections suggested by the \textit{Vocabulon}
algorithm.\looseness=-1

While we tested our methodology on the E. coli data, our approach is
potentially applicable to many other organisms, allowing researchers to
start to explore many other transcription networks such as those of
humans. In particular, there are many organisms for which far less of the
TRN structure is known a priori, making it impossible to use the
algorithms in \cite{liao1} and \cite{sabatti1}. In these cases our
$L_1$-penalization approach could still be applied provided an
``adequate'' prior could be generated. For example, in the case of human
data, one would probably rely on ChIP chip experiments to provide the
back-bone prior data on the possible location of binding sites. Finally,
it is worth recalling that, while we describe how to set the $\pi$ values
with specific reference to TRN, the $L_1$-penalized regression approach
can be used to estimate parameters of bipartite networks arising in other
scientific contexts.

\begin{appendix}\label{app}
\section*{Appendix: Identifiability}

\cite{liao1} provide the following sufficient conditions for identifiability
of the transcription regulation network model (\ref{basemodel}):
\begin{enumerate}
\item The connectivity matrix, $A$, must have full-column rank.
\item When a node in the regulatory layer is removed along with
all of the output nodes connected to it, the resulting network
must be characterized by a connectivity matrix that still has
full-column rank. This condition implies that each column of $A$
must have at least $L-1$ zeros.
\item$P$ must have full row rank. In other words, each
regulatory signal cannot be expressed as a linear combination of
the other regulatory signals.
\end{enumerate}

In our case these conditions were not satisfied because $L>T$ so $P$ was
not of full rank. However, the prior was very sparse with many zero
elements and relatively few values close to one, so it seemed
reasonable to
assume that the model was identifiable. To ensure this was correct, we ran
our fitting procedure $200$ times on the E. coli data, using randomized
starting values, and examined the resulting estimates for $P$.
Figure~\ref{bestworst} plots the best $20$ (left) and worst $20$ results
(right), in terms of the final objective values. There are some minor
differences in the estimates, but overall the results are encouragingly
similar.
This
experiment provided two useful pieces of information. First, it strongly
suggested that, at least for our prior, there were no identifiability
problems. Second, it also implied that the fitting algorithm was not
getting stuck in any local minima's and was reaching a global
optimum.
\end{appendix}

\section*{Acknowledgments}
The authors would like to thank the editor and two referees for helpful
comments and suggestions.

\clearpage
\renewcommand{\thefigure}{A.1}
%
\begin{sidewaysfigure}
\vspace*{20pt}
\includegraphics{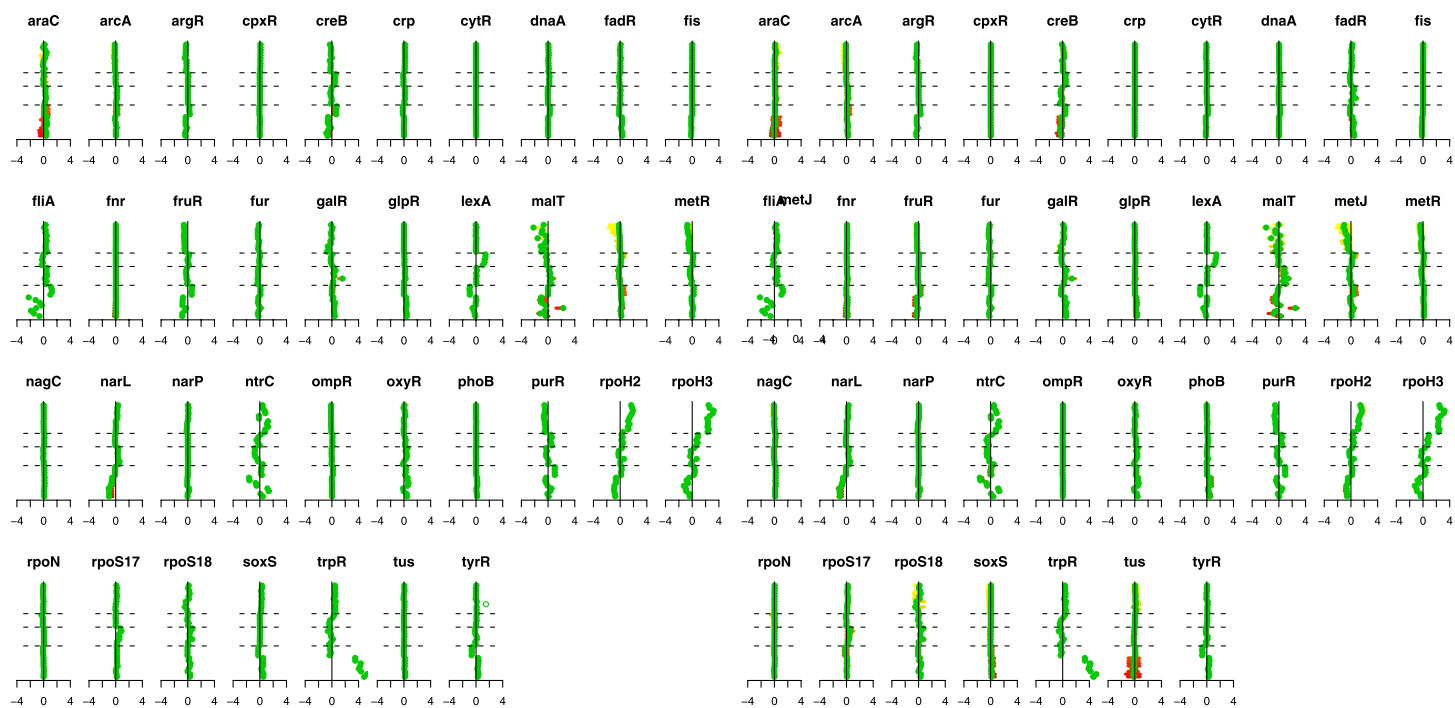}

\caption{Left: Best 20 runs. Right: Worst 20 runs.}\label{bestworst}
\end{sidewaysfigure}

\printaddresses

\end{document}